\documentclass[journal,letterpaper]{IEEEtran}

\usepackage{graphicx,cite,epsfig,amssymb,amsmath,amsfonts,multicol,subfigure,mathtools,bm,mathrsfs,setspace}
\usepackage{multirow}

\newcommand{\tabincell}[2]{\begin{tabular}{@{}#1@{}}#2\end{tabular}}

\begin{document}
\title{Cocktail Intra-Symbol-Codes: Exceeding the Channel Limit of QPSK Input}

\author{
	Bingli~Jiao,
	Mingxi~Yin
	and~Yuli~Yang
	\thanks{B. Jiao ({\em corresponding author}) and M. Yin are with the Department of Electronics and Peking University-Princeton University Joint Laboratory of Advanced Communications Research, Peking University, Beijing 100871, China (email: jiaobl@pku.edu.cn, yinmx@pku.edu.cn).}
	\thanks{Y. Yang is with the Department of Electronic and Electrical Engineering, University of Chester, Chester CH2 4NU, U.K. (e-mail: y.yang@chester.ac.uk).}
}

\maketitle

\begin{abstract}

This paper presents a new method, referred to as the cocktail intra-symbol-code, to reveal the possibility of exceeding the channel limit of QPSK input by transmitting independent signals in parallel manner and separating them the receiver, repetitively. In the proposed modulation scheme, the rhombic constellation is created to work with the modulated signal and the channel code in Euclidean- and Hamming space cooperatively. The theoretical derivations are done under the assumption of using the capacity-achieving codes and the results are shown in terms of the reliable bit rates. 
       	
\end{abstract}

\begin{IEEEkeywords}
reliable bit rate, channel capacity, Rhombic constellation, mutual information. 
\end{IEEEkeywords}

\IEEEpeerreviewmaketitle

\section{Introduction}
A significant underlying principle of the communication using the finite alphabet can be found at the additive white Gaussian noise (AWGN) channel model    
\begin{eqnarray}
\begin{array}{l}\label{equ1}
y = x + n,
\end{array}
\end{eqnarray} 
where $x$ is a finite alphabet signal, $y$ is the received signal and  $n$ is the random noise variable from a normally distributed ensemble of power $\sigma_N^2$, denoted by $n \sim \mathcal{N}(0,\sigma_N^2)$. 

The highest bit rate at which information can be transmitted with arbitrary small probability of error is up-bounded by \cite{Shannon1948} \cite{Verdu2011}
\begin{equation}
\begin{array}{l}\label{equ2}
{\rm{I}}(X;Y) ={\rm{H}}(Y) - {\rm{H}}(N)=\tilde{\rm{I}}_x(\gamma),
\end{array}
\end{equation}
where $\rm{I(X;Y)}$ is the mutual information, ${\rm{H}}(Y)$ is the entropy of the received signal and ${\rm{H}}(N) = {\log _2} (\sqrt{2 \pi e \sigma_N^2})$ is the entropy of the AWGN, and $\tilde{\rm{I}}_x(\gamma)$ is the mutual information expressed with  symbol energy to that of noise ratio, $\gamma=E_s/\sigma_N^2$, as the argument.  The result of \eqref{equ2} has been regarded as a law of the channel limit.  

However, for exceeding this limit, there were still some considerations, e.g., on a mathematical incentive  
\begin{equation}\label{equ-split}
\tilde{\rm{I}}_x(\gamma) <\tilde{\rm{I}}_{x_1}(\gamma_1) +\tilde{\rm{I}}_{x_2}(\gamma_2),
\end{equation}
where $x = x_1 + x_2 $ is the signal of the parallel transmission, and $x_1$ and $x_2$ are two independent signals.  The inequality is true due to the down-concavity of the mutual information \cite{ jiao} and  \cite{math}.     

Nevertheless, the great difficulties can be encountered when we try to squeeze out some effects in relation to \eqref{equ-split} on the signal designs.

This work is motivated by the above inequality to separate the signals of parallel transmission with respect to Euclidean- and Hamming space.  Eventually, we achieved a gain in terms of the reliable bit rates in comparison that of QPSK input.  

As a preparation for introducing the proposed method, we recall the conventional coded modulation scheme of BPSK that starts from the information sequence of $K$ bits in vector presentation, ${\bm c}=[c_1,c_2,...,c_K]\in \{0,1\}^K$, which is encoded by the code matrix ${\bm G}=\{g_{ij}\}$ into the channel code of $N$ bits ${\bm v} =  [v_1,v_2,... v_N] \in \{0,1\}^N$.  The channel code is  mapped onto BPSK symbols ${\bm x} =  [x_1,x_2,...,x_N] $ for the channel realization. Each symbol is drawn from a discrete constellation $S=\{+\alpha,-\alpha\}$, i.e., $x_i \in S$ and  $i=1,...,N$.  In general, the number of possible independent source vectors and that of the channel codewords are equal to $2^K$ with $R=K/N <1$, where $R$ is the code rate.  

Since the up-bound of the reliable communication bit rate is of interest in this research, we assume that there exist the capacity-achieving channel codes that allow the error-free transmission along with the above procedures of the demodulation and decoding.  This theoretical assumption is reasonable, because that the practical coded modulation of BPSK plus LDPC can approach the channel limit at a small gap of 0.0045dB \cite{LDPC}.

At the receiver, in the detection of each received symbol,  the estimate of $\hat{y} > 0$ or $\hat{y}< 0 $ provides an initial value to the channel decoder which chooses the codeword having the highest probability of being transmitted to recover the source codes. 

It is noted that the procedures of QPSK input are same as that of BPSK in principle.  Thus, we don't go through QPSK in the introduction.   

Throughout the paper, the lowercase bold letters denote vectors, uppercase bold letters denote matrices, e.g., ${\bm p}=[p_1,p_2, .... p_N]$. The operation of mo-2 plus of two vectors are expressed by $ {\bm p} \oplus {\bm q} = [p_1\oplus q_1, p_2 \oplus q_2, ...., p_N\oplus q_N] $, and the mutual information is expressed as  $\tilde{\rm{I}}(\gamma)$ with SNR as the argument. In addition, we use $\hat{z}$ to represent the estimate of the transmit quantity $z$ at the receiver.

\section{Communication Scheme}
In our theoretical approach, the derivations are done based on the assumption of error free transmissions of BPSK- and QPSK input.  When the Euclidean distance between any two constellation points is set equal to or larger than $2\alpha$, the up-bounds of the reliable bits rate can be calculated by    
\begin{equation}\label{assumtion-1}
\mathbb{R}^b=\hat{\rm{I}}_b(\alpha^2/\sigma_N^2),
\end{equation}
and 
\begin{equation}\label{assumtion-2}
\mathbb{R}^q=\hat{\rm{I}}_q(2\alpha^2/\sigma_N^2),
\end{equation}
where $\mathbb{R}^b$ and $\mathbb{R}^q$ are the reliable bit rates of BPSK and QPSK and $\hat{\rm{I}}_b$ and $\hat{\rm{I}}_q$ are the mutual informations, respectively.  Actually, we assume that there exist the capacity-achieving codes to work with the up-bounds.  

Let us consider a binary information bit sequence which is partitioned into three independent subsequences expressed in vector forms of ${\bm{c}}^{(1)}$, ${\bm{c}}^{(2)}$ and  ${\bm{c}}^{(3)}$, respectively.  

Using the capacity achieving code matrix to  ${\bm{c}}^{(1)}$, ${\bm{c}}^{(2)}$ yields 
\begin{equation}\label{code-matrix}
v^{(i)}_m = \sum\limits_{n = 1}^K {{g_{m,n}}  c^{(i)}_n}\ ,\ \ \ {\rm for}\ \ m=1,2,..,N,\ {\rm and} \ i=1,2
\end{equation} 
where ${\bm G}=\{g_{m,n}\}$ is the capacity-achieve encoder matrix,  $v^{(i)}_m$ is $m$-th competent of codeword ${\bm{v}}^{(i)}$, $K$ and $N$ are length of the information subsequences and that of the corresponding channel-codeword, respectively.

While, ${\bm{c}}^{(3)}$ is encoded by a different capacity-achieving code ${\bm{v}}^{(3)}$.  

To increase the energy efficiency of the proposed method,  we create a rhombic constellation with four possible points in Euclidean space as shown in Fig.1.  For simplicity, we refer this symbol to as the rhombic symbol $x^{rh,(k)}$.  To be specified, the rhombic symbol is expressed by $x^{rh,(1)} = 0+j\sqrt{3}\alpha $, $x^{rh,(2)} = \alpha+j0$,  $x^{rh,(3)}=0-j\sqrt{3}\alpha$ and $x^{rh,(4)} = -\alpha+j0$, where $j=\sqrt{-1}$.  

The Euclidean distance between two adjacent points in the constellation is $2\alpha$ so that the error free transmission applies as long as that of QPSK holds.  It is noted that the average symbol-energy is also same as that of QPSK, i.e., $E_s=2\alpha^2$.   

\begin{figure}[!t]
	\centering
	\includegraphics[width=0.3\textwidth]{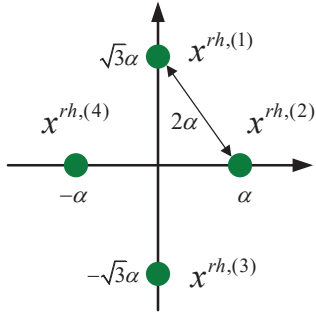}
	\caption{Constellations of the rhombic symbol.}
	\label{fig1}
\end{figure}

Before the signal modulation, the transmitter calculates ${\bm{v}}^{(1)} \oplus {\bm{v}}^{(2)} = {\bm{v}}^{(p)}$ and recodes each index $m'$, of which $v^{(p)}_{m'} = 1 $.  Then, the signal transmission is performed in the following two steps. 

The first step uses the rhombic symbol by mapping $v^{(p)}_m=1$ onto $x^{rh,(1)}_m$ or $x^{rh,(3)}_m$, and $v^{(p)}_{m}=0$ onto  $x^{rh,(2)}_{m}$ or $x^{rh,(4)}_m$.  The selection of $x^{rh,(1)}_m$ or $x^{rh,(3)}_m$ is determined by the component of $\bm{V}^{(3)}$: when $v^{(3)}_{l}=0$ uses  $x^{rh,(1)}_m$ and $v^{(3)}_{l}=1$ uses $x^{rh,(3)}_m$ for $l=1,2,..L$.  While, the selection of $x^{rh,(2)}_m$ or $x^{rh,(4)}_m$ is determined by the component of $\bm{V}^{(1)}$: when $v^{(1)}_m=0$ uses  $x^{rh,(2)}_m$ and $v^{(1)}_m=1$ uses $x^{rh,(4)}_m$.  

The signal modulation results are listed in Table I, where we use the subscript $m'$ to indicate the case of $v^{(p)}_{m'}=1$ and $m$ the case of $v^{(p)}_m=0$.
  
\begin{table}[!t]
	\renewcommand{\arraystretch}{1.5}
	\centering
	\small
	\caption{Signal modulation results.}
	\label{Table1}
	\begin{tabular}{c|c|c|c|c}
		\hline
		\tabincell{c}{$v^{(p)}_{m'}$}&\tabincell{c}{$v^{(1)}_{m'}$} & \tabincell{c}{$v^{(2)}_{m'}$} & \tabincell{c}{$v^{(3)}_{m'}$} &\tabincell{c}{Modulation Symbol} \\
		\hline		
		
		\tabincell{c}{$1$} & \tabincell{c}{$0$} & \tabincell{c}{$1$} & \tabincell{c}{$0$} & \tabincell{c}{$x^{rh,(1)}_{m'}$} \\
		\hline
		\tabincell{c}{$1$} & \tabincell{c}{$1$} & \tabincell{c}{$0$} & \tabincell{c}{$1$} & \tabincell{c}{$x^{rh,(3)}_{m'}$} \\
		\hline
		\hline
		\tabincell{c}{$v^{(p)}_{m}$} & \tabincell{c}{$v^{(1)}_{m}$} & \tabincell{c}{$v^{(2)}_{m}$} & \tabincell{c}{$-$} &\tabincell{c}{Modulation Symbol} \\
		\hline
		\tabincell{c}{$0$} & \tabincell{c}{$0$} & \tabincell{c}{$0$} & \tabincell{c}{$-$} & \tabincell{c}{$x^{rh,(2)}_{m}$} \\
		\hline
		\tabincell{c}{$0$} & \tabincell{c}{$1$} & \tabincell{c}{$1$} &\tabincell{c}{$-$} & \tabincell{c}{$x^{rh,(4)}_{m}$} \\

		\hline
		
	\end{tabular}
\end{table}

In the second step, transmitter transmits $v^{(1)}_{m'}$ by using BPSK modulation of $s_{m'} \in \{\alpha, -\alpha\}$, whereby  $v^{(1)}_{m'}=0$ is mapped onto $s_{m'}= \alpha$ or $v^{(1)}_{m'}=1$ onto $s_{m'}= -\alpha$.  The transmission is performed selecting components of subscripts $m'$ in sequence manner from $0$ to $N$, whenever $v^{(p)}_{m'}=1$ is found. The number of code components of $\{m'|v^{(p)}_{m'}=1\}$ is $N/2$ in average.  

To save the time resource, we layer two adjacent BPSK symbols perpendicularly into one QPSK symbol.  Thus, the second step transmission requires $N/4$ symbol durations to complete the transmission.  

We refer the proposed method to as the cocktail intra-symbol codes (CISC) method because the transmitted signals are composed of the different amplitudes of the independent signals in connections with their channel codes essentially.        

At the receiver, the signals of the first step transmission, i.e., the rhombic symbols, are demodulated and decoded for recovering $\bm{c}^{(p)}$ by taking $\hat{y}_m = x^{rh,(1)}_m$ or $x^{rh,(3)}_m$ for $v^{(p)}_m = 1$, and take $\hat{y}_m = x^{rh,(2)}_m$ or $x^{rh,(4)}_m$ for $v^{(p)}_{m} = 0$. 

The decoding results are ${\bm{c}}^{(p)}={\bm{c}}^{(1)} \oplus {\bm{c}}^{(2)}$, because of the property of the linear codes.  

Because of the error free transmission, we can use $\hat{\bm{c}}^{(p)}= {\bm{c}}^{(p)}$ to re-construct ${\bm{v}}^{(p)}$ without any error by
\begin{equation}\label{va}
\hat{v}^{(p)}_{m} = \sum\limits_{n = 1}^K {{g_{m,n}}  {\hat c}^{(p)}_n}\ ,\ \ \ {\rm for}\ \ m=1,2,..,N,
\end{equation}
where $\hat{v}^{(p)}_{m}$ is the re-constructed channel code component with $\hat{v}^{(p)}_{m}=v^{(p)}_{m} $.  

By calculating $\hat{\bm{v}}^{(1)} \oplus \hat{\bm{v}}^{(2)} = \hat{\bm{v}}^{(p)}$, the transmitter can find $m'$, of which $v^{(p)}_{m'} = 1 $, which is used to the second step transmission.  

\begin{figure}[ht]
	\centering
	\subfigure[]{
		\includegraphics[width=0.3\textwidth]{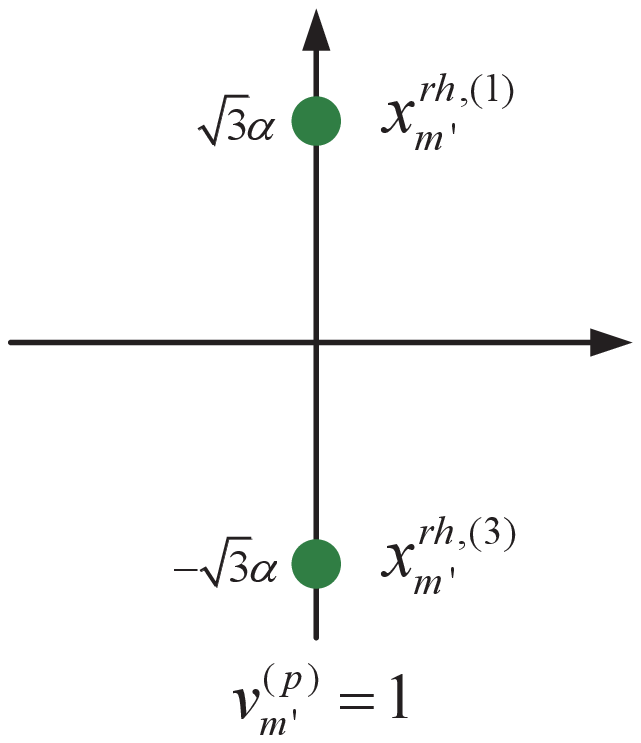}
		\label{fig2a}}
	\subfigure[]{
		\includegraphics[width=0.3\textwidth]{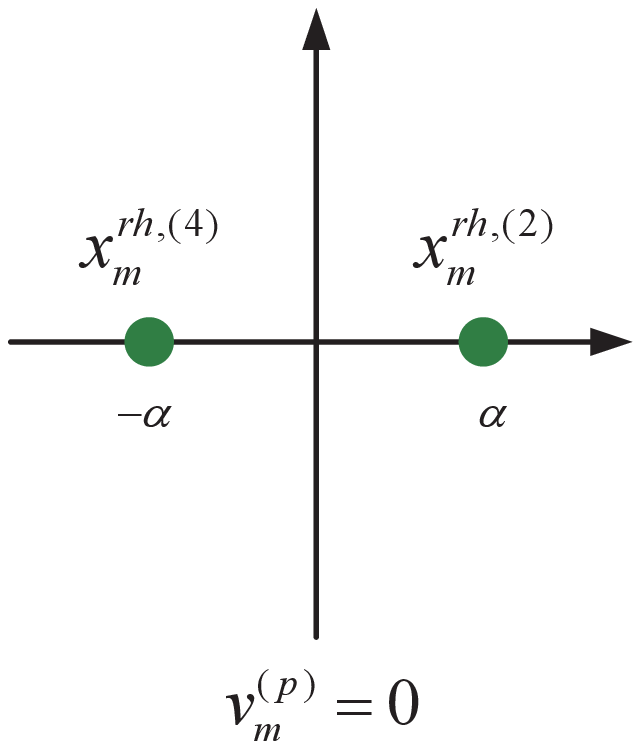}
		\label{fig2b}}
	\caption{Constellations of the $x^{rh}_{m}$ with (a) $v^{(p)}_{m'}=1$ (b)  $v^{(p)}_{m}=0$.}
	\label{fig2}
\end{figure}

Then the constellation of the $x^{rh}_{m}$ can be decoupled in Euclidean space: when $\hat{v}^{(p)}_{m'}=1$, the signal represents the BPSK modulation of $v^{(3)}_l$ along the vertical axis, and when $\hat{v}^{(p)}_{m}=0$, the signal represents the BPSK modulation of $v^{(1)}_m$ with respect to the horizontal axis as shown in Fig. \ref{fig2a} and \ref{fig2b}, respectively.  

We use the BPSK $x^{rh}_{m'}$ along vertical axis to the recovery of ${\bm{c}}^{(3)}$ by $\hat{y}_{m'}=0+j\sqrt{3}\alpha$ for $v^{(3)}_l=0$ and  $\hat{y}_{m'}=0-j\sqrt{3}\alpha $ for $v^{(3)}_l=1$, respectively.  Hence, the $\bm{c}^{(3)}$ can be fully recovered.   

Now, we use the rhpmbic symbols and the signals of the second step transmission the complete signal set for the recovering ${\bm{c}}^{(1)}$ as follows. The receiver uses the signals of the second step transmission and converts each QPSK symbol back to the original BPSK symbol and inserts each of them among the signals of the first step transmission, of which the BPSK is along with the horizontal axis, at the corresponding position of $v^{(p)}_{m'}=0$.  The conversion from QPSK to BPSK above does not suffer from any SNR loss, because both the signal energy and that of the noise are slit by a half.  

By demodulating and decoding over the complete signal set of the combined BPSK in connecting $\bm{v}^{(1)}$, we obtain $\hat{\bm{c}}^{(1)}$.   
  
Finally, $\hat{\bm{c}}^{(2)}$ can be obtained by 
\begin{equation}\label{code}
\hat{\bm{c}}^{(2)}=\hat {\bm{c}}^{(p)} \oplus \hat{\bm{c}}^{(1)}.
\end{equation}
where $\hat{\bm{c}}^{(1)}={\bm{c}}^{(1)}$ and $\hat{\bm{c}}^{(2)}={\bm{c}}^{(2)}$ can be guaranteed by the assumption of the error-free transmission based on the Euclidean distance of $2\alpha$.  
 
\section{Theoretic Contribution}

In this section, we compare the reliable bit rate of the CISC method with that of QPSK input. 

Let us examine $N_T= N+ N/4$ symbol durations, where $N_T$ is the total number of the transmitted symbols, $N$ is the number of the transmitted symbols of the first step and $N/4$ is that of the second. 

\begin{figure}[ht]
	\centering
	\includegraphics[width=0.49\textwidth]{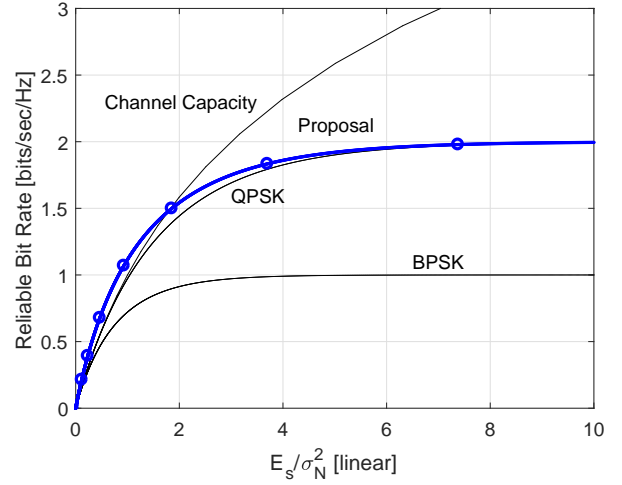}
	\caption{ADRs of cocktail BPSK compared with the channel capacity and BPSK versus linear ratio of $E_s/\sigma_N^2$ by \eqref{eqRt1}.}
	\label{fig4}
\end{figure}

\begin{figure}[!t]
	\centering
	\includegraphics[width=0.49\textwidth]{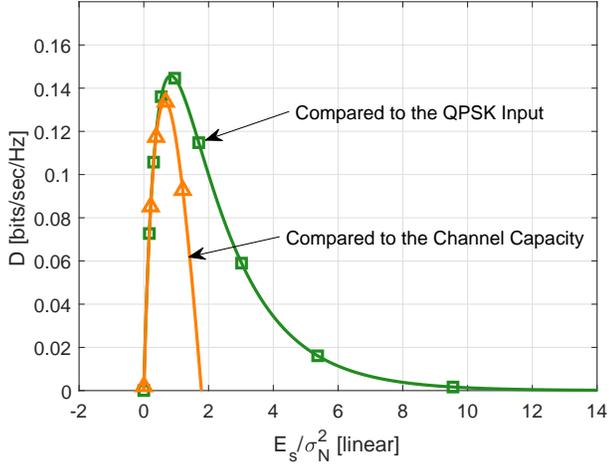}
	\caption{The difference between the proposed scheme and QPSK.}
	\label{fig5}
\end{figure}

The summation bits' number of ${\bm{c}}^{(1)}$ and ${\bm{c}}^{(2)}$ is $2K$.  Thus, the contribution of their bit rate over $N_T$ symbol durations can be averaged by 
\begin{equation}\label{code}
\mathbb{R}^q = \frac{N}{N_T}\tilde{\rm{I}}^q(2\alpha^2/\sigma_N^2) = \frac{4}{5} \tilde{\rm{I}}^q(2\alpha^2/\sigma_N^2),
\end{equation}
where $\mathbb{R}^q$ is the reliable bit rate of QPSK input, $\tilde{\rm{I}}^q(2\alpha^2/\sigma_N^2)$ is the mutual information of QPSK.

The bit rate of ${\bm{c}}^{(3)}$ can be calculated by  
\begin{equation}\label{code-rate}
\mathbb{R}^b=\frac{N/2}{N_T}\tilde{\rm{I}}^b(3\alpha^2/\sigma_N^2) = \frac{2}{5} \tilde{\rm{I}}^b(3\alpha^2/\sigma_N^2),
\end{equation}
where $\mathbb{R}^b$ is the reliable bit rate owning to the transmission of ${\bm{c}}^{(3)}$ and $\tilde{\rm{I}}^b(3\alpha^2/\sigma_N^2)$ is the mutual information of BPSK.

The total reliable bit rate of CISC method is found by
\begin{equation}\label{eqRt1}
\mathbb{R}^T(2\alpha^2/\sigma_N^2) = \frac{4}{5} \tilde{\rm{I}}^q(2\alpha^2/\sigma_N^2) + \frac{2}{5} \tilde{\rm{I}}^b(3\alpha^2/\sigma_N^2),
\end{equation}
where $\mathbb{R}^T$ is the reliable bit-rate of the proposed method.  

The numerical results of \eqref{eqRt1} are plotted in Fig.3, whereat one can find clearly that the reliable bit rate of CISC approach is larger than that of QPSK input over full range of SNR.

An analytic comparison can be made with the QPSK input at very low SNR, i.e.  $\gamma \to 0$ by
\begin{equation}\label{eqRt2}
\mathop {\lim }\limits_{2\alpha^2/\sigma_N^2  \to 0} \mathbb{R}^T(2\alpha^2/\sigma_N^2) = \frac{7}{5}(2\alpha^2/\sigma_N^2)\log_2 e,
\end{equation}
which is larger than $(2\alpha^2/\sigma_N^2)\log_2 e$ of the QPSK, as explained in the Appendix.

To show the spectral efficiency gain of CISC method, the numerical results of
\begin{equation}\label{eqD}
\mathbb{D} =\mathbb{R}^T(2\alpha^2/\sigma_N^2) - I^q(2\alpha^2/\sigma_N^2) ,
\end{equation}
are plotted as shown in Fig. \ref{fig5}, whereat one can find clearly the phenomenon of exceeding the spectral efficiency of QPSK input.

\section{Conclusion} 
In this paper, the new method called the cocktail intra-symbol-code is introduced to work in Hamming- and Euclidean Space for separating the parallel transmission of three information sources.  The higher reliable bit rates are obtained from the signal separations in Euclidean space with help of the channel codes ${\bm{V}}^{(p)}$ on the created rhombic symbol.  The results are found better than that of QPSK input.

\section*{Acknowledgment}   
This work was supported in part by the National Natural Science Foundation of China under Grant 61531004 and the Beijing Municipal Natural Science Foundation under Grant L172010.

\appendix

The first-order approximation of the QPSK using Taylor expansion is
\begin{equation}\label{f}
f(x) \approx f(0) + (df/dx)x,  
\end{equation} 
where $f(x)$ is the first-order approximation around the region of $x=0$.  

Because of $\hat{\rm{I}}_q(0)=0$, using \eqref{f} to calculate the first order-approximation of QPSK yields 
\begin{equation}
\hat{\rm{I}}_q(\gamma)\approx(d\hat{\rm{I}}_q/d\gamma|_{\gamma=0})\gamma  ,
\end{equation} 
and that of the channel capacity as
\begin{equation}
C(\gamma) \approx (dC/d\gamma|_{\gamma=0})\gamma = (\log_2e)\gamma  ,
\end{equation} 
where $\hat{\rm{I}}_q(\gamma)$ and $C(\gamma)$ represent the first-order approximation of mutual information of QPSK and that of channel capacity. 

According to Theorem 1 in~\cite{Verdu2011}, we obtain
\begin{equation}
\hat{\rm{I}}_q(\gamma)=C(\gamma) \approx (\log_2e)\gamma,  
\end{equation} 
when $\gamma << 1$.   

\end{document}